\documentstyle[preprint,aps,version2]{revtex}
\begin{document}
\draft
\preprint{\vbox{Submitted to Physical Review {\bf D}\hfill
	UC/NPL-1112}}
%
%
\tolerance=10000
\hbadness=10000
\tighten

\begin{title}
\centerline{Monte Carlo and Renormalization Group}  
\centerline{Effective Potentials in Scalar Field Theories}
\end{title}
\author{J.~R.~Shepard and V.~Dmitra\v sinovi\' c}
\begin{instit}Department of Physics, University of Colorado, 
Boulder CO 80309-0446
\end{instit}
\author{J.~A.~McNeil}
\begin{instit}
Department of Physics, Colorado School of Mines,
Golden CO 80401
\end{instit}
\begin{abstract}
We study constraint effective potentials for various strongly interacting $\phi^4$ 
theories.   Renormalization group (RG) equations for these quantities are discussed and
a heuristic development of a commonly used RG approximation is presented which stresses
the relationships among the loop expansion, the Schwinger-Dyson method and the renormalization
group approach.  We extend the standard RG treatment to account explicitly for finite
lattice effects.  Constraint effective potentials are then evaluated using Monte Carlo (MC) 
techniques and careful comparisons are made with RG calculations.  Explicit treatment of
finite lattice effects is found to be essential in achieving quantitative agreement with the 
MC effective potentials.  Excellent agreement is demonstrated for $d=3$ and $d=4$, $O(1)$ and 
$O(2)$ cases in both symmetric and broken phases.  
 
\end{abstract}
%
\newpage
\narrowtext
%
\section{INTRODUCTION}
 
Strong-coupled relativistic quantum field theories (RQFT's) play an essential role in any 
modern understanding of subatomic phenomena from the Higgs mechanism to fundamental theories
of nucleon and nuclear phenomena. There are important theoretical
issues, such as spontaneous symmetry breaking,  which can be studied with relatively 
simple scalar field theories.  Unfortunately,	
even the simplest interacting relativistic quantum field theories
are notoriously difficult to solve.  A classic example is the
issue of triviality in 4-dimensional $\phi^4$ scalar field theory which
has been tackled over the years using nearly the full complement of 
weapons in the theorists' arsenal (loop expansion,  strong coupling- and 
$\epsilon$-expansions, truncated systems of Schwinger-Dyson (SD) equations,
renormalization group (RG) techniques and Monte Carlo studies), 
{\it without} yielding a definitive 
answer. (See, {\it e.g.}, Ref.\cite{steph} .)  It is self-evident that
accurate and relatively simple approximate methods for solving RQFT's 
would be of enormous value.

In the present
paper we discuss approximations to $\phi^4$ theories emphasizing
throughout comparison with  lattice Monte Carlo calculations which, 
in a sense to be made explicit below, we consider exact.  In particular we will focus on 
the uniform properties which can be expressed in terms of ``effective potentials'' 
(or more precisely, ``constraint effective 
potentials'' as defined, {\it e.g.}, in Ref.\cite{oraif} ) from which,
for example, zero-momentum $n$-point couplings may be extracted.  The 
present results may be summarized most succinctly by stating that the
renormalization group (RG) effective potential -- evaluated so as to 
account for lattice artifacts -- provides a spectacularly accurate
description of its lattice counterpart with little calculational
complexity for a wide variety of scalar field theories including symmetric
and broken phases in 2, 3, and 4-D with O(1) and O(2) symmetries.
In contrast, other methods such as the use of truncated SD 
equations\cite{rivers} yield only qualitative agreement at best.  Moreover, as will be 
emphasized below, the simple RG effective potential used here is known
to be the first term in a momentum expansion meaning that systematic 
improvements in the approximation are possible.  Perhaps even more 
important is the evidence that -- in contrast to, {\it e.g.}, the 
hierarchy of SD equations -- this expansion appears to be strongly convergent
for cases with strong coupling.  Indeed, estimates of next-to-leading 
order correction to be made below are quite small but consistently 
improve the already extraordinary agreement with lattice results.
Finally, as will also be addressed below, the RG approach is ideally 
suited to the construction of {\it effective} low energy theories which
can economically account for the long-wavelength behavior of the full
theories.  We will also discuss some anticipated exploitations of this 
possibility.
 
In section II we review briefly the history of the version of the RG effective potential 
used in the present work providing
highlights and references to several other recent studies. In section III we provide
a heuristic derivation of the RG equation for the effective potential and
compare with truncated Schwinger-Dyson.  In section IV we describe the methods we use in 
extracting the effective potential from Monte Carlo simulations and in solving the 
RG equations numerically. In section V we present the results of our calculations
comparing Monte Carlo and RG descriptions for broken and symmetric phases of O(1) and 
O(2) $\phi^4$ field theories in  3- and 4-dimensions.

\section{REVIEW OF THE RG EFFECTIVE POTENTIAL}

	We now trace the development of the form of the RG equations
used in the present study.  The original work by Wilson on the renormalization
group\cite{wilsona} was in essence a discussion of how to obtain the effective action at 
long wavelengths from a bare action at short wavelengths.  He and his coworkers
discussed a variety of approximations to the full (and intractable) RG equations.
Shortly thereafter, Wegner and Houghton\cite{wandh} derived an exact RG equation
for the effective potential by integrating over thin shells in momentum space, 
a technique referred to as the ``sharp cutoff method.''  (See Ref.\cite{wegner} for
an early discussion of this and alternative ``smooth cutoff'' methods.  Also see
Ref.\cite{weinberg} for early discussions of many RG concepts.  Ref.\cite{polchinski}
provides a later discussion of some of these ideas in the course of constructing
a novel and intuitively appealing RG-based proof of the renormalizability of
$\phi^4$ field theories.)   A seminal early application of the Wegner and
Houghton method to $\phi^4$ theories was presented by Hasenfratz and 
Hasenfratz\cite{handh} .  (See also Ref.\cite{handn} \ for an
illuminating comparison of analyses of the cutoff dependence of
the Higgs mass using the Wilson-based and Wegner/Houghton-based
RG methods.)  These authors obtained a tractable approximation to the exact 
Wegner and Houghton 
equation by projecting the full effective action onto an effective potential
for which all $n$-point couplings are momentum independent.  They
used the resulting RG equation to extract information about RG flows
in the vicinity of Wilson and Gaussian fixed points in $d=3$ and to establish 
the absence of a Wilson fixed point in $d=4$.  They also extracted various 
critical exponents for the $d=3$ case and found them to be in good agreement 
with values determined by a variety of other techniques.

	There has been a recent resurgence of interest in RG effective
potentials.  Wetterich and coworkers\cite{tandw} , for example, have
extended the approach of Hasenfratz and Hasenfratz using the so-called 
``smooth cutoff'' method in deriving their RG equations.  Their technique
permits evaluation of, {\it e.g.}, the anomalous dimension, $\eta$, which
is implicitly zero in the Hasenfratz and Hasenfratz approximation and yields
values of critical exponents which are in impressive agreement with
those determined by other methods.  (See Ref.\cite{golner} for an early
calculation of $\eta$ using an extension of one of Wilson's original approximate
RG equations.)  This increase in calculational power
comes, however, at the cost of a substantial increase in numerical 
complexity.  In other work, Morris\cite{morrisa} \ gives an in-depth
discussion of the RG effective potential as derived using both sharp and
smooth cutoffs. He also\cite{morrisb} developes a version of the 
systematic momentum expansion alluded to above and presents an analysis
of $\phi^4$ in $d=3$ using the first correction to what is in effect the 
Hasenfratz and Hasenfratz method (which appears as the zero-momentum limit 
in both the sharp- and smooth-cutoff derivations).  Without attempting to 
be comprehensive, we mention other related papers which appear here as
Refs.\cite{kuti} ,\cite{other} and \cite{landg}.

\section{A HEURISTIC DEVELOPMENT OF THE RG EQUATION FOR THE
	EFFECTIVE POTENTIAL}

	Derivations of RG equations for the effective action or their 
reduction to expressions governing the effective potential 
appear in Refs.\cite{wilsona,wandh,polchinski,handh,tandw,morrisa,landg}  
and will not be reproduced here.  Instead we provide a heuristic 
and, we hope, intuitive development of the RG equation in the 
zero-momentum limit which has as its point of departure the familiar
expression for the one-loop contribution to the effective action of
a uniform system.  

	In $d$ Euclidean dimensions, the bare action for a one-component
$\phi^4$ theory is given by
\begin{equation}
	S_0=\int\ d^dx\ \biggl[{1\over 2}(\partial_\mu\phi)^2
	+ U_0(\phi)\biggr] \label{ca}
\end{equation}
where
\begin{equation}
	U_0(\phi)={1\over 2} m_0^2 \phi^2 + {1\over 4} \lambda_0 \phi^4 .
	\label{cb}
\end{equation}
The one-loop perturbative correction to the bare potential for a 
uniform system corresponds to the familiar expression (primes denote
differentiation with respect to $\phi$)
\begin{equation}
	U_1(\phi)={1\over 2} \int{{d^dk}\over{(2\pi)^d}}\ 
	\ln\bigl[k^2+U_0^{''}(\phi)\bigr]
	\label{cc}
\end{equation}
which can be interpreted physically as the sum of zero-point energies
of vacuum modes for a free scalar meson theory with meson 
mass-squared $m^2=U_0^{''}(\phi)$.
This quantity is divergent and we regularize it by adopting a UV cutoff 
$\Lambda_0$: 
\begin{equation}
	U_1(\phi)=
	{1\over 2} \int{{d^dk}\over{(2\pi)^d}}\ 
	\theta(\Lambda_0-|\vec k|)\ 
	\ln\bigl[k^2+U_0^{''}(\phi)\bigr]
	\label{cd}
\end{equation}
We note that the lowest order 
truncated Schwinger-Dyson equation can be obtained from the following expression for 
what we will call the Schwinger-Dyson effective potential, $U_{SD}$, in which the 1-loop 
contribution $U_1(\phi)$ as given above is self-consistent in the following sense:
\begin{equation}
	U_{SD}(\phi)\simeq U_0(\phi)+{1\over 2} \int{{d^dk}\over{(2\pi)^d}}\ 
           \theta(\Lambda_0-|\vec k|)\ 
	   \ln\bigl[k^2+U_{SD}^{''}(\phi)\bigr].
	   \label{ce}
\end{equation}

       The RG equation for the effective potential $U_{RG}$ may be obtained heuristically
by starting with the loop expansion of the effective action, the first term of which is 
given in Eq.~\ref{cd}.  Assume we know $U(\phi,\Lambda)$, the effective potential at 
momentum scale $\Lambda$.  We then have the following expression for 
$U(\phi,\Lambda-d\Lambda)$, the effective potential at a slightly smaller momentum scale:
\begin{eqnarray}
 U(\phi,\Lambda -d\Lambda)&=& U(\phi,\Lambda) 
	 +{1\over 2} \int{{d^dk}\over{(2\pi)^d}}\ \theta(\Lambda-|\vec k|)\ 
	  \ \theta(|\vec k|-\Lambda+d\Lambda)\ \ln\bigl[k^2+U_0^{''}(\phi)\bigr] \cr
         &=&U(\phi,\Lambda)+{A_d\over 2} \Lambda^{d-1} 
	  \ln\big[\Lambda^2+U''(\phi,\Lambda)\big]\ d\Lambda
         +O(d\Lambda^2)
\end{eqnarray}
where $A_d=\int d\Omega_d/(2\pi)^d=1/(2^{d-1} \pi^{d/2} \Gamma(d/2)\ )$ and 
the higher order loops contribute at $O(d\Lambda^2)$. Taylor expanding the left 
hand side and taking the limit $d\Lambda\rightarrow 0$ gives the RG equation:
\begin{equation}
	{{dU_{RG}(\phi,\Lambda)}\over{d\Lambda}}=
	-{{A_d}\over 2}\ \Lambda^{d-1}\ 
	\ln\bigl[\Lambda^2+U_{RG}^{''}(\phi,\Lambda)\bigr].
	\label{ch}
\end{equation}
The essential point is that where the Eq.~\ref{ce} requires $U_{SD}$ be made 
self-consistent across all momentum scales at once, Eq.~\ref{ch} ensures that 
$U_{RG}$ is self-consistent at every scale.  Put another way, the evolution (or 
``flow'') of $U_{RG}$ at some scale $\Lambda$ is determined self-consistently 
by $U_{RG}$ at that same scale.  We make think of $U_{SD}$ as the result of the 
crudest possible discretization of the flow which yields $U_{RG}$, namely one in 
which the integration of the differential equation in Eq.~\ref{ch} is carried out
in a single step.

The extension to $O(N)$ is straight-forward.  Defining $\rho\equiv\sqrt{\vec\phi^2}$
and with primes denoting differentiation with respect to $\rho$, we find
\begin{equation}
	{{dU_{RG}(\rho,\Lambda)}\over{d\Lambda}}=
	-{{A_d}\over 2}\ \Lambda^{d-1}\ \biggl\{ \ln\bigl[\Lambda^2
		+U_{RG}^{''}(\rho,\Lambda)\bigr]
	+(N-1) \ln\bigl[\Lambda^2+U_{RG}^{'}(\rho,\Lambda)/\rho\bigr]\biggr\} .
	\label{ci}
\end{equation}
To make contact with, {\it e.g.}, Hasenfratz and Hasenfratz\cite{handh}, 
we scale all dimensionful quantities by the appropriate power of $\Lambda$ so 
as to render them dimensionless.  For example,  
$\phi\rightarrow x\equiv\phi/\Lambda^{d/2 - 1}$ (where in this approximation 
the anomalous dimension, $\eta$, vanishes).  Designating the rescaled potential as
\begin{equation}
	\tilde U_{RG}(x,t)\equiv U_{RG}(\phi,\Lambda)/\Lambda^d ,
	\label{cii}
\end{equation}
where $t\equiv\ln(\Lambda_0/\Lambda)$, the one-component equation (Eq.~\ref{ch}) 
becomes, after dropping irrelevant constants, becomes 
\begin{eqnarray}
	-\Lambda{{d\tilde U_{RG}(x,\Lambda)}\over{d\Lambda}}&=&
	{{d\tilde U_{RG}(x,t)}\over{dt}}= \nonumber\\
	&&{{A_d}\over 2} \ln\bigl[1+\tilde U_{RG}^{''}(x,t)\bigr]
	+(1-d/2)\ x\ \tilde U_{RG}^{'}(x,t)+d\ \tilde U_{RG}(x,t)
	\label{cj}
\end{eqnarray}
where primes  now denote differentiation with respect to $x$. This is 
Eq.~22 of Ref.~\cite{handh} (with $\eta=0$).   Finally, defining 
$f(x,t)\equiv\tilde U_{RG}^{'}(x,t)$, Eq.~\ref{cj} becomes  
\begin{equation}
	\dot f={{A_d}\over 2}\ {{f^{''}}\over{1+f^{'}}}
	\ +\ (1-d/2)\ x f^{'}\ +\ (1+d/2)\ f
	\label{ck}
\end{equation}
where the ``dot'' means differentiation with respect to $t$.  This is 
Eq.~9 of Ref.~\cite{handh} and is a form of the RG equation for the 
effective potential frequently encountered in the literature.

	These RG equations may be integrated to find the effective 
potential at momentum scale $\Lambda$ subject to the {\it boundary
condition} that the bare and effective potentials coincide at the
UV cutoff scale $\Lambda_0$.  In so doing, quantum corrections associated 
with single meson modes with momenta between $\Lambda$ and $\Lambda_0$
are (approximately) included in the effective potential while 
contributions from modes at lower frequencies must still be treated 
explicitly.  As touched upon above, the power of the RG approach derives from 
its self-consistency {\it at each momentum scale}.  This feature is evident in 
Eq.~\ref{ch} where it is clear that the evolution of $U_{RG}$ at momentum scale 
$\Lambda$ depends on a quantum correction which is consistent with the 
potential itself at that particular momentum scale.  

Another important and perhaps surprising aspect of the RG equations is
that -- while they have the {\it form} of a truncated loop expansion -- no 
such truncation has been made.  As originally shown by Wegner and 
Houghton\cite{wandh} 
higher-order loops are suppressed by powers of the 
differential $d\Lambda$ and their equation therefore has the form of a one-loop 
result but is in fact {\it exact}.  Thus the RG equation for the effective potential, 
obtained by Hasenfratz and Hasenfratz as an approximation to the Wegner and Houghton 
expression has nothing to do with explicitly 
dropping high order quantum corrections but instead depends on projecting 
the full effective action onto a form with independent $n$-particle 
couplings.  (This is what is meant by the ``zero-momentum'' RG effective
potential.)   Of course, just as is the case when solving truncated SD 
equations\cite{rivers}, quantum corrections of arbitrarily high order in a
perturbative loop expansion are automatically generated in the course of
integrating the RG equation.  
(The relationship between  the RG equation 
and the loop expansion was touched upon by Polchinski\cite{polchinski} and 
addressed in more detail by Morris\cite{morrisa} who also made a direct 
comparisons with the SD expression given below in  Eq.~\ref{ceb}.)  

Once the effective potential is obtained, the renormalized mass and 
4-point coupling are given by:
\begin{equation}
m^2={d^2 U_{eff}\over d\phi^2}\bigg|_{\phi \rightarrow \phi_0}
\end{equation}
\begin{equation}
\lambda ={1\over 6} {d^4 U_{eff}\over d\phi^4}\bigg|_{\phi \rightarrow \phi_0}
\end{equation}
where $\phi_0$ is the value of the field at the minimum of $U_{eff}$.  For example 
in the case of the Schwinger-Dyson equation, Eq.~\ref{ce}, in the symmetric phase  
where $\phi_0=0$,  we have
\begin{equation}
	m_{SD}^2=m_0^2 + 
	3\lambda_{SD}\int {{d^dk}\over{(2\pi)^d}}\ {1\over{k^2+m_{SD}^2}}
	\label{cea}
\end{equation}
where we have used $U_{SD}^{''}(\phi=0)\leftrightarrow m_{SD}^2$ and 
$U_{SD}^{''''}(\phi=0)\leftrightarrow 6\lambda_{SD}$. Approximating 
$\lambda_{SD}$ by $\lambda_0$ yields
the familiar (see, {\it e.g.}, Eq.~2.17 of Ref.\cite{rivers}) 
SD self-consistency relation for the ``dressed'' mass:
\begin{equation}
	m_{SD}^2=m_0^2+3\lambda_0\int 
	{{d^dk}\over{(2\pi)^d}}\ {1\over{k^2+m_{SD}^2}}
	\label{ceb}
\end{equation}
where the self-consistent nature of the SD method is apparent. 
(We note that the approximation defined by this equation for $m_{SD}$ is 
also known\cite{morrisa} as the ``cactus'' approximation.)  Likewise, within the same 
approximation, the 4-point is
\begin{equation}
	\lambda_{SD}=\lambda_0/\biggl[1+9\lambda_0\ 
	\int{{d^dk}\over{(2\pi)^d}}\ {1\over{(k^2+m_{SD}^2})^2} 
	\biggr] .\label{cec}
\end{equation}

Examination of RG renormalized masses and 4-point couplings following from 
Eq.~\ref{ch} (or the equivalent expressions given subsequently) 
requires finding numerical solutions to the non-linear partial differential RG 
equations. In the next section we discuss our numerical methods for solving the RG 
equations and for determining the ``exact'' effective potentials from our Monte 
Carlo calculations.

\section{Numerical Methods}

We have carried out RG calculations for the effective constraint potential
in two complementary ways. In the first, appropriate to studying the 
continuum limit, we treat $\Lambda$ as a continuous variable. 
In the second approach, appropriate
to direct comparison with our Monte Carlo calculations which of course refer to a finite 
lattice, we treat the lattice normal modes explicitly.  We now give detailed descriptions of 
both methods.

	In the first approach, we solve Eq.~\ref{ch} by treating 
$\Lambda$ as a continuous variable. (We have also evaluated the equivalent expressions, 
Eq.~\ref{cj} or Eq.~\ref{ck}. Although one 
version is often better suited to a given calculation than the others, we 
found them to be entirely equivalent to one another as expected.)
We begin by expanding $U_{RG}(\phi,\Lambda)$ in a power series in $\phi$ truncated 
at $\phi^{2M}$:
\begin{equation}
	U_{RG}(\phi,\Lambda)=\sum_{i=1}^M\ {1\over{2i}}\ 
	{\cal U}_{2i}(\Lambda)\phi^{2i} .
	\label{da}
\end{equation}
The boundary condition referred to in Section III is imposed by setting
\begin{equation}
	{\cal U}_2(\Lambda_0)=m_0^2,\quad {\cal U}_4(\Lambda_0)=\lambda_0,
	\quad
	{\rm and}\quad {\cal U}_{2i}(\Lambda_0)=0\quad {\rm for}\quad
	2i\ge 6 .
	\label{db}
\end{equation}
The range of the $\Lambda$ integration is divided into $N$ intervals of equal length 
$\Delta\Lambda$:
\begin{equation}
         \Lambda_n = \Lambda_0 - n \Delta\Lambda
	\label{dc}
\end{equation}
with
\begin{equation}
	\Delta\Lambda= (\Lambda_0 - \Lambda_{IR})/N,
\end{equation}
where $\Lambda_{IR}=2\pi /N_{lat}$ is the infrared scale 
appropriate to a lattice of $N_{lat}^d$ sites and $\Lambda_0$ is 
chosen to yield the same phase space volume as the finite sum-over-modes.

A similar decomposition is made of a range of $\phi$, the other independent
variable:
\begin{equation}
	\phi_0 < \ldots \phi_j < \phi_{j+1} < \ldots< \phi_J .
	\label{dd}
\end{equation}
Each step in the integration is accomplished by first making a least squares fit
of the polynomial on the r.h.s of Eq.~\ref{da} to $U_{RG}(\phi,\Lambda_n)$
in order to determine the ${\cal U}_{2i}(\Lambda_n)$ and then computing
\begin{equation}
	U_{RG}(\phi_j,\Lambda_{n+1})=U_{RG}(\phi_j,\Lambda_n)+
	\Delta\Lambda\cdot\Lambda_n^{d-1}\ {{A_d}\over 2}\ 
	\ln\bigl[\Lambda_n^2+U_{RG}^{''}(\phi_j,\Lambda_n)\bigr]
	\label{de}
\end{equation}
for each $j$ 
where $U_{RG}^{''}(\phi_j,\Lambda_n)$ is computed analytically from the
polynomial fit.  Starting with $n=0$, this procedure is iterated until 
ultimately the
$J+1$ values of $U_{RG}(\phi_j,\Lambda_{IR})$ are found.  We refer to this 
as the ``continuum RG effective potential'' (C-RG).   Obviously this
method yields directly the $m$-point couplings, 
${\cal U}_{m}(\Lambda_{IR})$, as well.  
The calculation can sometimes be unstable in the vicinity of the 
$\phi$-endpoint, $\phi_J$, and some care is required in choosing values of 
$M$, $N$ and $J$ to ensure stable integration.  In the calculations to be 
presented below, we typically used
\begin{equation}
	M\simeq 10,\quad N\simeq 4000,\quad {\rm and}\quad J\simeq 50 .
	\label{dea}
\end{equation}

	In our second approach, an alternate RG calculation was formulated to 
facilitate comparison with the lattice results to be described below.  The derivative 
terms in the action (Eq.~\ref{ca}) are approximated on the lattice by a 
finite difference formula.  For a uniform system on a $N_{lat}^d$ lattice 
with periodic boundary conditions, the vacuum modes are momentum 
eigenstates with wavenumbers in each direction given by 
$\kappa_n\equiv 2n\pi/N_{lat}$, $n=1,2,\ldots N_{lat}$ where we take the lattice constant 
to be unity, $a=1$.  We make the following identification:
\begin{equation}
	\int{{d^dk}\over{(2\pi)^d}}\ f(k^2)\rightarrow
	{1\over{N_{lat}^d}}\ \sum_{n_1=1}^{N_{lat}}\ 
	\sum_{n_2=1}^{N_{lat}}\ldots \sum_{n_d=1}^{N_{lat}}\ 
	f(k_{n_1}^2+k_{n_2}^2+\ldots+k_{n_d}^2)
	\label{df}
\end{equation}
where $k_n^2\equiv4\sin^2(\kappa_n/2)$.  Now define the $N_{lat}^d$ 
scalar quantities
\begin{equation}
	k_{n_1 \ldots n_d}^2\equiv
	k_{n_1}^2+k_{n_2}^2+\ldots+k_{n_d}^2 ,
	\label{dg}
\end{equation}
then sort them by magnitude and relabel:
\begin{equation}
	k_{1}^2\ge\ldots k_{{\cal N}}^2\ge k_{{\cal N}+1}^2\ge
	\ldots\ge k_{N_{lat}^d}^2
	\label{dh}
\end{equation}
where $k_{1}^2 = 4 d$ and $k_{N_{lat}^d}^2=0$.
Our ``latticized'' version of the RG equation for the effective potential, Eq.~\ref{de}, is
\begin{equation}
	U_{RG}(\phi_i,k_{{\cal N}+1}^2)=U_{RG}(\phi_i,k_{{\cal N}}^2)
	+{1\over{N_{lat}^d}}\ {{A_d}\over 2}\  
	\ln\bigl[k_{{\cal N}}^2+U_{RG}^{''}(\phi_i,k_{{\cal N}}^2)
	\bigr]
	\label{di}
\end{equation}
with the boundary condition
\begin{equation}
	U_{RG}(\phi_i,k_{1}^2)=U_0(\phi_i) .
	\label{dj}
\end{equation}
We will refer to this as the ``lattice RG effective potential'' (L-RG). 

	The lattice Monte Carlo calculations to be presented below employed 
$d$-dimensional cubic lattices with periodic boundary conditions in conjunction 
with the Metropolis algorithm to sample the $\phi$-configurations with weights given by 
the exponentiated action.  The Monte Carlo observable of primary interest here
is the  ``constraint effective potential''\cite{oraif} which we refer to as the
``MC effective potential'' and designate by $U_{eff}$.  Formally, this quantity
is defined via 
\begin{equation}
	\exp\bigl[-N_{lat}^d U_{eff}(\bar\phi)\bigr]=
	\int{\cal D}\{\phi\}\ \delta(\bar\phi-\sum_i\phi_i/
	N_{lat}^d)\ \exp\bigl[-S[\phi]\bigr]
	\label{dk}
\end{equation}
where $\sum_i$ means a sum over all lattice sites and $\bar\phi$
is the {\it average} field on the lattice
for a given field configuration in the path integral.  

Since the Monte Carlo method generates sequences of field configurations,
$[\phi]_n$, weighted by $\exp\bigl[-S[\phi]_n\bigr]$, $U_{eff}
(\bar\phi)$ is extracted from the lattice Monte Carlo calculation from a
histogram of the number of configurations generated {\it vs} the
average value of the field for those configurations.  Let 
$d{\cal N}(\bar\phi)$ be the number of configurations with average
field values in an interval $d\bar\phi$ about $\bar\phi$.  Then 
\begin{equation}
	d{\cal N}(\bar\phi)\propto 
	\exp\bigl[-N_{lat}^d U_{eff}(\bar\phi)\bigr]\ d\bar\phi
	\label{dla}
\end{equation}
and, to within an overall constant,
\begin{equation}
	U_{eff}(\bar\phi)=-{1\over{N_{lat}^d}}\ 
	\ln {{d{\cal N}(\bar\phi)}\over{d\bar\phi}} .
	\label{dl}
\end{equation}
For $O(N)$ theories this expression is modified slightly.  Let
$\bar\rho$ be the average value of $\rho\equiv\sqrt(\vec\phi^2)$
on the lattice and let $d{\cal N}(\bar\rho)$ be the number of 
configurations in an interval $d\bar\rho$ about $\bar\rho$.
Then, since
\begin{equation}
	d{\cal N}(\bar\rho)\propto\ \bar\rho^{N-1}\ 
	\exp\bigl[-N_{lat}^d\ U_{eff}(\bar\rho)\bigr]\ d\bar\rho ,
	\label{dm}
\end{equation}
we have
\begin{equation}
	U_{eff}(\bar\rho)=-{1\over{N_{lat}^d}}\ 
	\ln\biggl[{1\over{\bar\rho^{N-1}}}
	{{d{\cal N}(\bar\rho)}\over{d\bar\rho}}\biggr] .
	\label{dn}
\end{equation}
Of course the most probable average field values are near the minimum
of $U_{eff}$.  In order to determine $U_{eff}$ away from its minimum;
{\it i.e.}, to sample a range of relatively improbable values of 
$\bar\phi$ (or $\bar\rho$), it is often necessary to add a constraint
to the updating algorithm which rejects ``moves'' which take the
average field value outside a preset ``window''\cite{window}.  By
matching $U_{eff}$'s for a number of adjacent windows, it is possible
to detemine the lattice effective potential over a wide range of 
average field values.  (See Ref.~\cite{tsypin} for an interesting 
alternative method.)

	Finally we comment on the ``convexity'' of $U_{eff}$.
It can be shown by very general arguments\cite{oraif} that
the condition $U_{eff}^{''}(\bar\phi)\ge 0$ (primes
denote differentiation with respect to $\bar\phi$) must be satisfied for
all $\bar\phi$.  Presently, we will show $U_{eff}$'s for ``broken''
symmetry phases of $\phi^4$ theories which {\it do not} satisfy the
convexity condition.  The apparent contradiction is resolved by noting 
that convexity is required only in the thermodynamic limit; {\it i.e.},
as $N_{lat}\rightarrow\infty$.  For the finite lattice results to be
presented below, convexity is not a fundamental property of $U_{eff}$.

\section{Comparison of MC and RG Effective Potentials}

	We now present MC effective potentials as defined in 
Eqs.~\ref{dl} or \ref{dn} for a variety of cases.
We also show the corresponding continuum and lattice RG effective 
potentials of Eqs.~\ref{de} and \ref{di}, respectively.  In computing
the former RG potential, we always set the IR cutoff; {\it i.e.}, 
$\Lambda_N$ in Eq.~\ref{dc}, to $2\pi/N_{lat}$.  Fixing the UV cutoff, 
$\Lambda_0$, is ambiguous.  We adopt the prescription that $\Lambda_0$ 
should be chosen so that the phase space volumes integrated over should be 
the same for continuum and lattice calculations.  This correspondance is 
made quantitative by setting $f(k^2)=1$ in Eq.~\ref{df} which implies,
in effect,
\begin{equation}
	A_d\ \int_{0}^{\Lambda_0}\ k^{d-1}\ dk=A_d\ 
	{{\Lambda_0^d}\over{d}}=1.
	\label{ea}
\end{equation}
Then, {\it e.g.},
\begin{eqnarray}
	\Lambda_0 (d=2) &=& (4\pi)^{1/2} \nonumber \\
	\Lambda_0 (d=3) &=& (6\pi^2)^{1/3} \label{eaa} \\
	\Lambda_0 (d=4) &=& (32\pi^4)^{1/4} . \nonumber
\end{eqnarray}
In the calculations shown in this work the bare coupling was fixed at $\lambda_0=10$ while 
the bare mass $m_0$ was adjusted to give a desired renormalized mass. The MC
effective potentials were fit with a polynomial of the form
\begin{equation}
	U_{eff}(\phi)={1\over 2}m_{fit}^2\phi^2 
	+ {1\over 4}\lambda_{fit}\phi^4
	\label{eb}
\end{equation}
to determine the renormalized MC masses and 4-point couplings.

In our RG calculations we also kept the 4-point coupling fixed at $\lambda_0=10$.
We now describe our procedure for determining the bare mass in the RG calculations. 
For symmetric phases, the RG bare masses were adjusted to yield
${\partial^2 U_{RG}}/{\partial\phi^2}|_{\phi=0}=m_{fit}^2$.  For 
broken phases, the RG bare masses were adjusted so as to give the best fit
``by eye'' to $U_{eff}$ in the vicinity of its minimum (see discussion below).
Since the MC and RG effective potentials in the symmetric phase behave identically 
at small $\phi$ {\it by construction}, we typically compare -- for both symmetric and 
broken phases -- MC and RG results for the subtracted quantity $U-{1\over 2}m_{fit}^2\phi^2$ 
rather than results for the $U$'s themselves.  Table 1 summarizes results of $O(1)$
and $O(2)$ calculations in both symmetric (S) and broken (B) phases in
$d=3$ and $d=4$.  For the RG results, $m\equiv\sqrt{{\cal U}_2}$ and 
$\lambda\equiv{\cal U}_4/6$.  The Table shows that our bare masses squared
are typically $m_0^2\sim -2^2$ while the renormalized masses squared are
$m^2\sim \pm 0.2^2$ to $\pm 0.3^2$ with renormalized couplings of 
$\lambda\sim 1$ to $4$.  

Clearly quantum corrections are large and simple perturbation theory ({\it e.g.}, the
loop expansion) 
cannot account for them.  For comparison purposes we examine results from the simplest 
Schwinger-Dyson (SD) equations\cite{rivers}.  Consider, for example, 
a ``latticized'' $d=3$, $O(1)$ version of Eq.~\ref{ceb}:
\begin{equation}
	m_{SD}^2=m_0^2+3\lambda_0\ {1\over{N^3_{lat}}}\ \sum_{{\cal N}=1}^{N^3_{lat}}
	{1\over{k^2_{\cal N}+m^2_{SD}}}
	\label{ec}
\end{equation}
where the $k^2_{\cal N}$'s are defined in Eq.~\ref{dh}.
In this expression, a value of $m^2_0=-2.61^2$ is required to give $m_{SD}=0.285$ 
for $\lambda_0=10$ and $N_{lat}=16$ while the corresponding MC calculation yielding
$m_{fit}=0.285$ requires $m^2_0=-2.16^2$.  The latticized expression for the SD 
4-point coupling, Eq.~\ref{cec}, is
\begin{equation}
	\lambda_{SD}=\lambda_0/
	\biggl[1+9\lambda_0\ {1\over{N^3_{lat}}}\ \sum_{{\cal N}=1}^{N^3_{lat}}
	{1\over{(k^2_{\cal N}+m^2_{SD})^2}}\biggr],
	\label{ed}
\end{equation}
and gives $\lambda_{SD}=0.65$ as compared to the MC value of $1.49$.  As another 
example, we find that, for $d=4$, $O(1)$, the MC-determined renormalized 
mass is $m_{fit}=0.276$ using $m_0^2= -1.88^2$ compared to the latticized SD value 
$m_0^2=-2.13^2$ needed to yield the same renormalized mass.  
Similarly, the latticized SD equation for the 4-point coupling gives 
$\lambda_{SD}=1.28$ {\it vs} $\lambda_{MC}=3.42$.   Not surprisingly, these simple 
SD equations do not in general provide a quantitatively accurate picture of the 
quantum effects.   Although we do not display SD results for the effective potential, 
it is apparent from the 4-point couplings that the SD-derived effective potential 
will yield at best only a qualitative description of the MC effective potential.

We turn now to the Monte Carlo calculations of the effective potential and comparisons 
with the two RG effective potentials, C-RG and L-RG, discussed in 
the previous section.  As mentioned previously, since we adjust the bare mass in the
RG calculations to give the MC fit mass, $m_{fit}$, in most of the figures that follow 
we display the subtracted quantity, $U-{1\over 2}m_{fit}^2\phi^2$, to emphasize 
the predictive power of our calculations.

Figure \ref{figb} displays the subtracted ({\it i.e.}, 
$U-{1\over 2}m_{fit}^2\phi^2$) MC effective potential for $d=3$, $O(1)$
in the symmetric phase on a $16^3$ lattice.   A MC bare mass squared
of $-2.16^2$ yields $m_{fit}=0.285$ and $\lambda_{fit}=1.49$, as mentioned
above and shown in Table 1.  Note that we have used the ``windowing''
technique described in Section IV to extend the range of $\phi$ at which 
the MC effective potential is determined from a maximum value of about
$0.1$ which is feasible without windowing to a maximum value of $0.22$.
(The relative normalizations of adjacent windows were determined by
matching intermediate values well determined in both windows.)
However, the values of $m_{fit}$ and $\lambda_{fit}$ were determined by fits 
to the MC values of $U_{eff}$ for $\phi\leq 0.1$ only.  To reproduce this value of 
$m_{fit}$, a C-RG bare mass squared of $-1.91^2$ was required.
Of course this numerical value depends on our prescription for $\Lambda_0$
which was discussed above.  (However, the C-RG effective potential
is only weakly dependent on $\Lambda_0$ provided the RG bare mass is adjusted 
to give a particular value of the renormalized  mass, $m$.)
The C-RG value for the effective 4-point coupling is $\lambda=1.89$ and
it is evident from Figure \ref{figb} that the C-RG calculation
(solid line) does {\it not} accurately reproduce the quartic and higher
$\phi$-dependence of the MC effective potential.  By comparison, a bare mass squared of
$-2.14^2$ was required in the L-RG calculation to give
$m=m_{fit}$.  That the L-RG bare mass differs from the MC value
by less than 1\% is especially gratifying; in contrast, the C-RG bare 
mass differs by about 12\% while the SD difference is also
about 12\% but in the opposite direction.  The L-RG value of 
$\lambda=1.43$ is also quite close to $\lambda_{fit}=1.49$.  However,
inspection of Figure \ref{figb} --- where the agreement between MC and
L-RG results over the wide range of $\phi$ spanned using the windowing 
technique is spectacular --- suggests that even this small discrepancy
is probably due to the fact that the lattice $U_{RG}$ contains terms of order
$\phi^6$ and higher while our fit to $U_{eff}$ does not.  We note that our limited MC
``data'' are not good enough to extract the $\phi^6$ or higher terms reliably.
Also appearing in Figure \ref{figb} as the dashed line is the L-RG result omitting
all contributions of order $\phi^6$ and higher.  Comparison of this curve with the MC
results show that these high order contributions are important for large values of 
$\phi$ and that the L-RG accounts for them quite well.  
To summarize, the L-RG calculation which includes lattice effects in the integration 
of the differential equation for $U_{RG}$ yields {\it excellent quantitative}
agreement with MC effective potentials even on a relatively small 
lattice provided a very small adjustment in the RG bare mass 
is permitted.   For the C-RG calculation, comparison with the MC results is not 
straightforward because of the ambiguity in choosing $\Lambda_0$ which then results in
ambigiuties in the bare mass.  Futhermore, even when the bare mass is allowed to vary 
freely, agreement is poorer than for the L-RG.

	Unsubtracted effective potentials for the $d=3$, $O(1)$ case in the 
{\it broken} phase appear in Figure \ref{figca}.  As mentioned above, the bare masses 
were adjusted in the RG calculations to yield the correct location of the minimum.  
From Figure \ref{figca} we see again that the description of the effective potential
provided by the L-RG calculation
is excellent.  The C-RG calculation, however, is incapable of 
simultaneously reproducing the location in $\phi$ and the depth of the minimum in 
the MC effective potential.
The fit shown is a (non-unique) compromise.
The subtracted potentials for this case are shown in Figure \ref{figc} and again 
demonstrate the superior description of $\phi^6$ and higher contributions 
to the effective potential provided by the L-RG.  Indeed, the description 
of the MC results by the L-RG is again spectacular.  The 
bare mass squared used in this calculation is $-2.23^2$ which differs
from the MC value of $-2.24^2$ by less than 1\%.  (The high quality of this
fit again leads us to attribute the differences between $\lambda_{fit}$ and 
$\lambda$ from the L-RG calculation --- 1.48 and 0.52, respectively --- to the
absence of $\phi^6$ and higher terms in the fitting function rather than to
shortcomings of the L-RG method.)

	Similar findings were obtained for $d=4$, $O(1)$ as well as $O(2)$ cases 
in $d=3$ and $d=4$ in both symmetric and broken phases.  Values of the bare and 
renormalized masses and 4-point couplings are reported in Table I. 
The L-RG description of the subtracted MC effective potential is nearly perfect 
in all cases and is distinctly superior to that provided by the C-RG.  We note 
that this is true even when the correlation length ({\it i.e.}, $m^{-1}$) is
large compared to the overall lattice size.  For our $d=4$, $O(2)$ case, a
small MC mass of $0.06$ was found ({\it i.e.}, the correlation length was 
$\sim 15$) on an $8^4$ lattice (!) and, yet, as shown in Figure \ref{figh}, the
L-RG calculation agrees almost prefectly with the subtracted MC effective
potential.  Furthermore, the L-RG bare masses never differ by more 
than 1.5\% from the MC values and are typically much closer.  Overall, we
conclude that the L-RG method provides an extraordinarily accurate,
semianalytic description of lattice MC calculations of the effective 
potential at a tiny fraction of the computational cost.   

\section{Summary and Conclusions}

	We have presented a discussion of the renormalization group (RG) effective
potential for $\phi^4$ scalar field theories which stresses its relationship to  
the loop expansion and the Schwinger-Dyson (SD) method.  We show in a simple way that 
the unique feature of the RG method -- which is presumably the origin of its ability 
to describe the full theory as well as it does -- is its self-consistency at
all momentum scales.  We have formulated a version of the RG equation for the 
effective potential which explicitly treats finite lattice effects so that
RG calculations may be compared directly with lattice Monte Carlo (MC) calculations 
without extrapolations of the latter to infinite lattice size and zero lattice spacing 
(See Ref.~\cite{tsypin} and references contained therein for some examples of such 
extrapolations).   We compare such ``latticized'' or L-RG potentials along with
standard continuum or C-RG potentials\cite{handh} to constraint effective potentials 
obtained via MC methods.  The L-RG effective potentials are in excellent quantitative 
agreement with the MC results for $d=3$ and $d=4$, $O(1)$ and $O(2)$ cases in 
both symmetric and broken phases.  More specifically, we find that, upon adjusting 
the L-RG bare masses so as to yield renormalized masses equal to those found in the MC 
calculations, the L-RG bare masses never differ by more than 1.5\% from the MC values 
and are typically much closer.  The renormalized 4-point couplings found by solving 
the L-RG equations are also typically in close agreement with the MC values.  In at 
least one case for which the MC effective potential was determined over a wide
range of $\phi$ using windowing techniques, there is strong evidence that $\phi^6$
and higher contributions are also well accounted for in the L-RG approach.  In
another case, we showed that the L-RG approach works very well even when the
correlation length is appreciably larger than the overall lattice size 
($m\ N_{lat}\simeq 0.5$).
Comparison of C-RG calculations with the MC results was found to be ambiguous; 
however, even when ambiguous parameters were tuned to give optimal agreement 
with the MC quantities, this accord was found to be inferior to that achieved
using the L-RG without tuning.  Finally, latticized SD equations were seen to 
yield only a rough qualitative description of of MC masses and 4-point couplings.

	We conclude that the L-RG method provides an very accurate, relatively simple 
description of lattice MC calculations.  It may prove to be a useful guide in
extrapolating finite lattice results to the continuum limit since the relation
between the L-RG and the C-RG is simple and well-defined.  We may also be able to 
apply it to some specific problems of current interest such as accounting for the
``Universal Effective Potential'' for $\phi^4$ field theories in $d=3$ recently
obtained via MC techniques by Tsypin\cite{tsypin}.  The success of the method
encourages us to make various extensions of it.  For example, we plan to go beyond
the ``zero-momentum'' approximation so as to be able to compute wavefunction
renormalizations\cite{golner,morrisc}.  We have also begun to examine including
Yukawa-coupled fermions in the approach which would be a crucial step toward
our long term goal of constructing effective theories of low-energy nuclear
phenomena whose field theoretic structure is fully understood and for which we 
have an accurate and relatively simple approximation scheme of proven reliability.

\section{Acknowledgements} The authors gratefully acknowledge valuable
discussions with Steven Pollock and Anna Hasenfratz.  This work supported in part 
by the U.S.D.O.E.

\vfill
\eject
\newpage
%
%
%
%

%
%
%
\newpage
\figure{ Subtracted effective potentials for the $d=3$, $O(1)$ case in
	the symmetric phase.  The solid dots are the MC results while the
	solid and dotted lines correspond to the C-RG and L-RG calculations,
	respectively.  Statistical uncertainties for the MC effective potential 
	are not shown but are typically smaller than the size of the dots.  
	The dashed curve is the L-RG result with $\phi^6$ and
	higher contributions omitted.  Input and output quantities are discussed 
	in the text and are presented in Table I. \label{figb}}
\figure{ Effective potentials for the $d=3$, $O(1)$ case in
	the broken phase.  The solid dots are the MC results while the
	solid and dotted lines correspond to the C-RG and L-RG calculations,
	respectively.  Input and output quantities are discussed 
	in the text and are presented in Table I. \label{figca}}
\figure{ Same as Figure \ref{figca} but for the subtracted effective potentials.
	\label{figc}}
\figure{ Same as for Figure \ref{figc} but for $d=4$, $O(2)$ in
	the symmetric phase. \label{figh}}

\begin{table}
\caption{Input and ouput quantities for Monte Carlo (MC) as well as continuum (C-RG) 
	and lattice (L-RG) renormalization group calculations for $d=3$ and $d=4$, 
	$O(1)$ and $O(2)$ cases in both symmetric (S) and broken (B) phases.  The 
	bare masses used are denoted by $m_0$ while the MC renormalized masses and 
	4-point couplings are $m_{fit}$ and $\lambda_{fit}$, respectively, as defined
	in the text and their RG counterparts are $m$ and $\lambda$.  The MC
	and L-RG calculations use the same value of $N_{lat}$.}
\begin{tabular}{ccccccccccccccc}
 & & &\multicolumn{4}{c}{MC}& &\multicolumn{3}{c}{C-RG}& &\multicolumn{3}{c}{L-RG} \\
\tableline
Case & Phase & & $N_{lat}$ & $m_0$ & $m_{fit}$ & $\lambda_{fit}$ & & $m_0$ & $m$ & 
	$\lambda$ & & $m_0$ & $m$ & $\lambda$ \\
\tableline
 & & & & & & & & & & & & & & \\
$d=3$, $O(1)$ & S & & 16 & 2.16$i$ & 0.285    & 1.49 & & 1.91$i$ & 0.285    & 1.89 & 
	 &2.14$i$ & 0.285    & 1.43 \\
              & B & & 16 & 2.24$i$ & 0.264$i$ & 1.48 & & 2.00$i$ & 0.258$i$ & 1.10 & 
	 &2.23$i$ & 0.227$i$ & 0.52 \\
 & & & & & & & & & & & & & & \\
$d=4$, $O(1)$ & S & &  8 & 1.88$i$ & 0.276    & 3.42 & & 1.62$i$ & 0.277    & 4.49 & 
	 &1.86$i$ & 0.276    & 3.47 \\
              & B & &  8 & 1.97$i$ & 0.388$i$ & 3.98 & & 1.74$i$ & 0.409$i$ & 4.40 & 
	 &1.94$i$ & 0.365$i$ & 3.00 \\
 & & & & & & & & & & & & & & \\
$d=3$, $O(2)$ & S & & 16 & 2.55$i$ & 0.251    & 1.34 & & 2.26$i$ & 0.251    & 1.67 & 
	 &2.55$i$ & 0.250    & 1.22 \\
              & B & & 16 & 2.62$i$ & 0.221$i$ & 1.17 & & 2.34$i$ & 0.229$i$ & 1.07 & 
	 &2.63$i$ & 0.199$i$ & 0.59 \\
 & & & & & & & & & & & & & & \\
$d=4$, $O(2)$ & S & &  8 & 2.22$i$ & 0.060    & 3.28 & & 1.92$i$ & 0.060    & 4.09 & 
	 &2.20$i$ & 0.060    & 3.10 \\
              & B & &  8 & 2.25$i$ & 0.256$i$ & 3.01 & & 1.96$i$ & 0.291$i$ & 4.32 & 
	 &2.23$i$ & 0.258$i$ & 2.93 \\
 & & & & & & & & & & & & & & \\
\end{tabular}
\end{table}

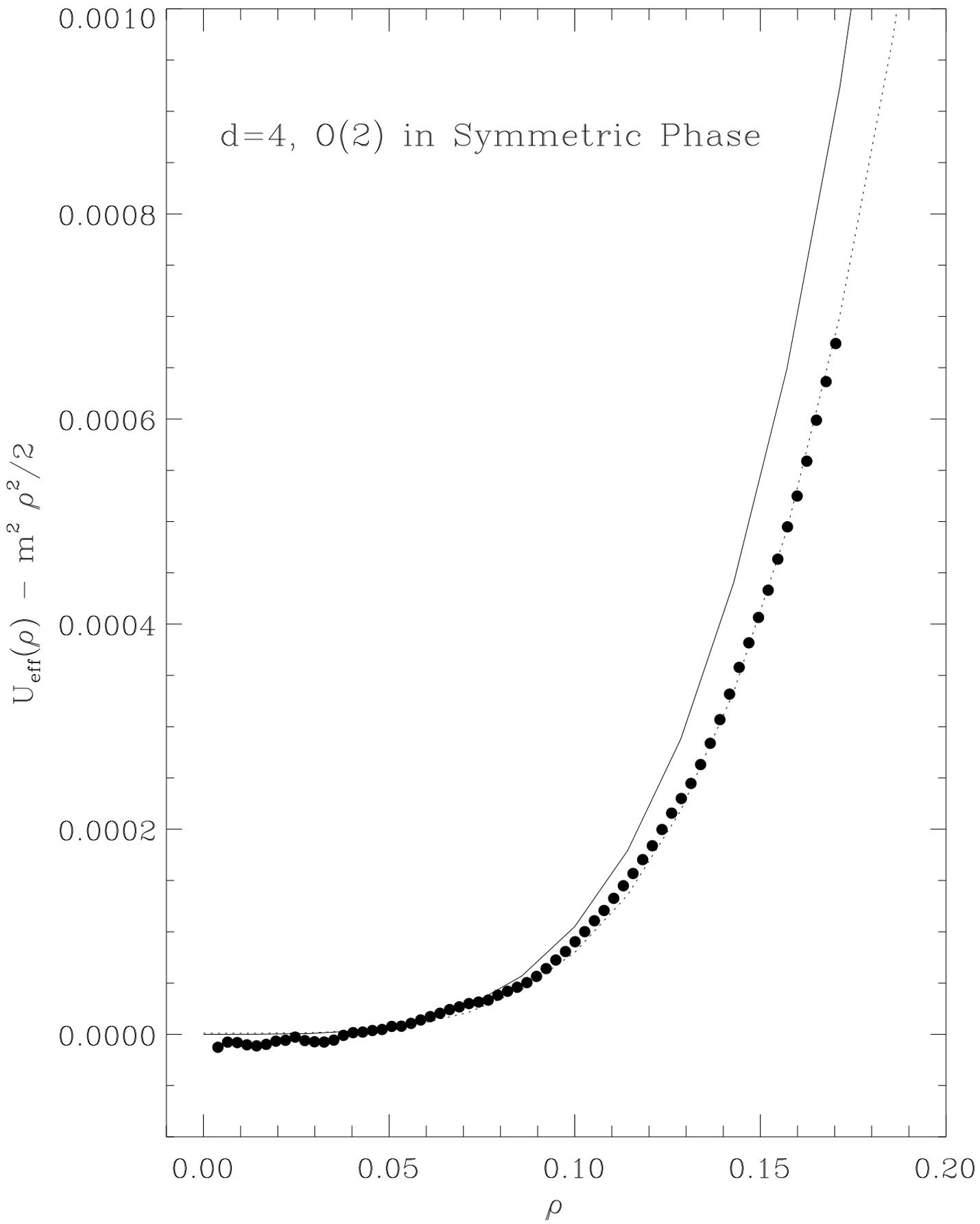
\end{document}